# Multi-cloak invisibility, a new strategy for simultaneous acoustic and electromagnetic invisibility

Hasanpour Tadi Saeed[1], shokri Babak[1, 2, *]

[1]Laser and Plasma research institute, Shahid Beheshti University, G.C., P.O. Box, 19839-69411, Tehran, Iran.

[2]Physics Department of Shahid Beheshti University, G.C., P.O. Box, 19839-69411, Tehran, Iran. (Email: b-shokri@sbu.ac.ir)

Based on electromagnetic and acoustic transformation theory, a new strategy has been presented in this article to implement double invisibility cloaking, which has not been done yet. By applying a combination of four conventional cloaking methods, for cloaking an object by electromagnetic and acoustic waves, which are two essential methods for objects detection (sonar, radar, eye, ear, camera ...), the object will be blind or deaf, while it will be undetectable acoustically and electromagnetically simultaneously. In deaf mode structure, the object has an interaction with electromagnetic waves; so it can see what is outside the cloaking structure electromagnetically while it has not any acoustic interaction with the outside, but in blind mode arrangement, as opposed to the deaf arrangement, the acoustic link has turned on while there is no electromagnetic connection. Finally, the proposed approach in this paper does not generalize for any two classes of physics, electrical-mechanical, thermal-mechanical, electrical-thermal, electromagnetic-sonic, etc.

**Keywords**: Cloaking, acoustic cloaking, electromagnetic cloaking, transformation optics, transformation acoustics, invisibility

## Introduction

Today, the ability to create hidden objects is an interesting subject to scientific researchers. After Pendry published his article about Transformation Optics (TO), invisibility has been no longer an imagination [1]. In addition to invisibility, there are many applications based on TO theory such as cloaking device [1-6], perfect lens [7-9], electromagnetic (EM) wave controlling [10-13], etc. After TO method proved its strength in electromagnetism, researchers in other branches of physics and engineering developed an interest in exploring and employing transformation coordination in acoustic [14-17], elastic waves [18], thermodynamics and heat transfer [19], seismic waves [20], fluid flow [21], etc. The most important studies in the application of transformation coordinates are fields and waves controlling, new device design, and invisibility. However, what gave us this method, which was not imaginable before, is invisibility.

Electromagnetic invisibility makes objects upon radar, eyes, camera, night vision camera, and any electromagnetic waves detector out of reach. Acoustic cloaking makes objects hidden against sonar, ears and any acoustic sensors and seismic waves cloaking can protect buildings against earthquake disturbed oscillations. In any branch of physics, there are some methods to create an ideal cloak, but nothing is perfect [22]. In any type of cloaking (EM, acoustic …), there are many techniques for detection, and the hidden object with perfect cloaking, is still detectable [23-25], although somewhat complicated. If an object is ideally cloaked against EM waves and cannot be detectable electromagnetically, it can be detected by other physical methods, for example by acoustic waves.

While invisibility is a fascinating subject, multi-cloak invisibility is super-fascinating. In the field of heat and electricity conduction, it has been done earlier [27]. Furthermore, there is an example of triple carpet cloaking to hide objects against acoustic, EM, and water waves [28]. In the governing equation and physical parameters (electrical and thermal conductivity of materials) of heat and electric conduction, there are many similarities. Hence, it is not difficult to discover materials which conduct electricity and heat (For example, copper is an electric and thermal conductor) and it will have the same behavior for heat and heat transfer. Heat and electric current are transferred, in addition to the material's conductivity depending on the structure of the geometric design. They determine the distribution of electrical and thermal flow. These features have been used in double thermal and electric current cloaking [27], with a single structure. In contrast to the similarity of the governing equations of electric and heat conduction, if this similarity does not exist (for example acoustic and EM wave propagation), it would not be possible to provide simultaneous cloaking by a single structure with a geometric design. In the field of invisibility or cloaking, inhomogeneous, anisotropic, and controllable materials are required. Fortunately, metamaterials have the power to create structures with appropriate properties. However, it is so difficult to build a metamaterial which works simultaneously in two or more branches of physics (electromagnetism, acoustics, thermodynamics ...). So we cannot produce a multi- cloaking shell with a single structure.

There is no physical object which is perfectly invisible, but if the detection method is more complicated, the object's detection possibility decreases. Perfect invisibility is a dream. We can get closer to it, but it never happens. Two most important methods for detection are based on acoustic and EM waves scattering. If an object is cloaked simultaneously by these two detection methods, it is very hard to track.

In this paper, we introduce a new procedure for cloaking an object by acoustic and EM waves simultaneously. This method presents two modes of cloaking, blind and deaf, which is based on a combination of perfect cloaking [2] and out of shell cloaking [5]. In blind mode, according to Fig.1, two cloaking structures, acoustic out of shell cloak (AOSC) and electromagnetic perfect cloak (EMPC), together make the object double-cloaked. EMPC structure cloaks any object positioned inside the structure and AOSC is an external device which hides objects outside the AOSC structure. By combining these two structures, a double-cloaking area can be created.

As it can be seen in fig.2, EM wave scattering simulation shows that EM waves cannot penetrate inside the EMPC shell and in the waveform of transmission waves are preserved. On the other hand, AOSC can hide objects located outside the structure acoustically as shown in fig.3, so if the object and EMPC are positioned near the AOCS, the surface can be cloaked from acoustic waves. For this purpose, two anti-images were embedded inside the complementary area (for the object and EMPC). This structure can cloak objects outside the shell, as shown in fig.3. In this arrangement, only the acoustic waves reach the object not EM waves, so this arrangement is called blind mode.

In the deaf mode, the structure is similar to the blind arrangement, with two differences; an outer shell (EMPC) with a perfect acoustic cloak (PAC) and an internal structure (AOSC) with EM waves out of shell (EMOSC) are replaced. This arrangement is shown in Fig.4. Due to the PAC layer, acoustic waves in the PAC are canceled as shown in Fig.5. While EMOSC makes the object invisible, it can interact with EM waves, as shown in fig.6. So this arrangement is called deaf mode.

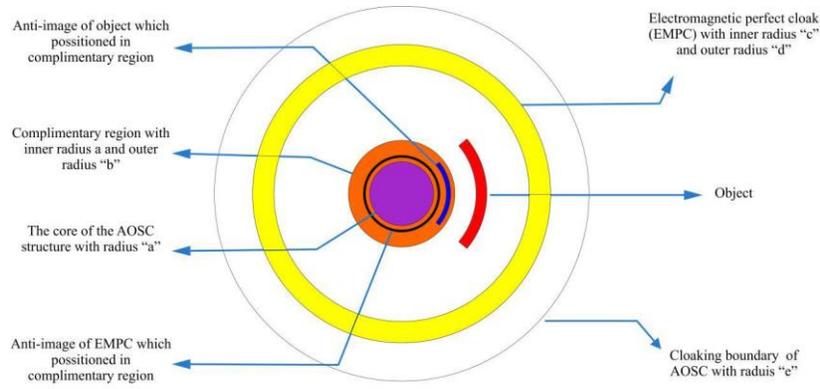

**Figure 1: blind arrangement to the double-cloaking.**

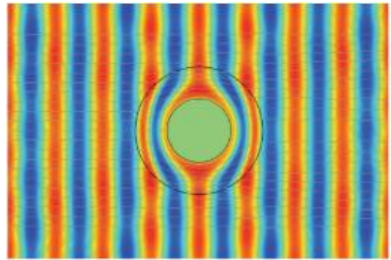

**Figure 2: pattern of electromagnetic plane wave interaction with perfect cloak**[2]**.**

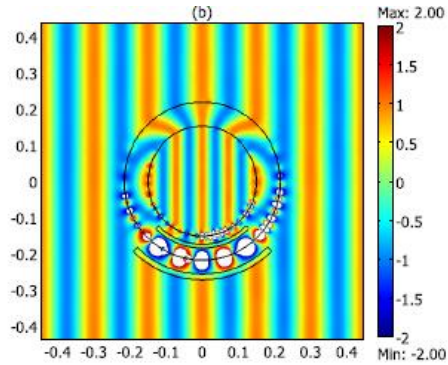

**Figure 3: out of shell acoustic cloaking**[17]**.**

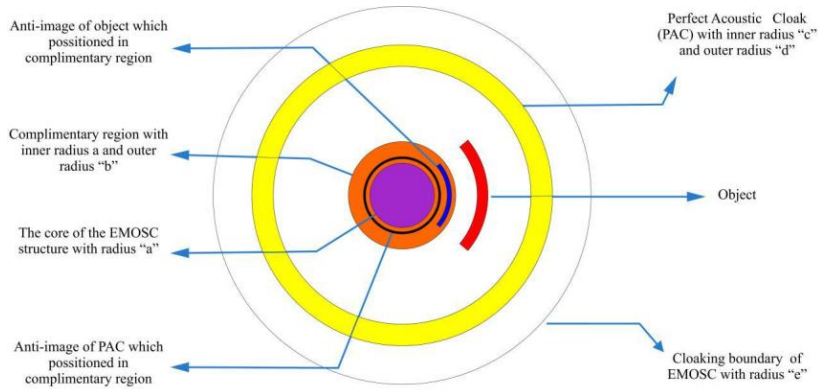

**Figure 4: deaf arrangement to the double-cloaking.**

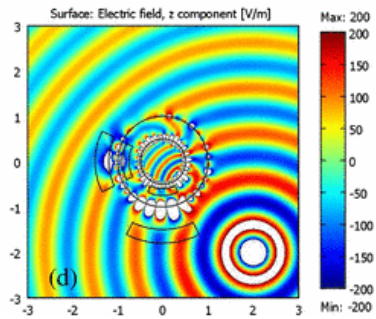

**Figure 5: electromagnetic out of shell cloaking**[5]**.**

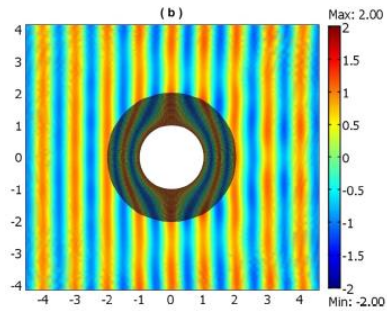

**Figure 6: perfect acoustic cloaking structure**[17]**.**

## Materials and methods:

In the perfect cloaking, the object does not interact with the EM waves, so there is no absorption, scattering, and dispersion for the waves. Hence, when a wave passes within the invisible body, its waveform and phase remain unchanged. As previously mentioned, an electromagnetically hidden object is apparent for other waves (acoustic, elastic ...) or detection methods. If a PAC shell covers an EMPC structure, they are acoustically invisible, but the PAC layer destroys the electromagnetic cloak because the EMPC can hide objects inside its construction and the PAC layer located outside. So the PAC layer is visible for EM waves and invisible for acoustic waves. Out of shell invisibility is a solution to this problem; the object is undetectable while interacting with the EM waves or acoustic waves. In this method, any target could be located outside the cloaking structure, so if another layer joined to the structure, this layer could be cloaked, the same as the target, while there is not any destructive influences on the invisibility. Thus, from the acoustic and EM points of view, the target and the added layer are cloaked. If the added layer is a PAC, then the object will be simultaneously cloaked, consequently we call this arrangement double-cloaking structure. This approach can be applied to other methods of cloaking objects. As previously discussed, two essential modes of invisibility, deaf and blind for simultaneous acoustic and EM cloaking, will be addressed in the following part.

**Blind cloak**

According to Fig.7, there are four regions: two regions for acoustic out of shell cloaking (I, II), an object region (III) and a perfect EMW cloak (IV). In this mode, a combination of acoustic out of shell cloak (AOSC) and perfect electromagnetic cloak (EMPC) have been used. In a 2D cylindrical AOSC, there are three essential areas: a core area with inner radius "a," which matches the wave phase, a complementary area located between spaces "a" and "b," which produced anti-image for outside until the radius "e." Objects inside this area are cloaked via creating an anti-image region embedded inside the complementary structure. AOSC material parameters can be obtained according to the presented method in Ref- [17]. Hence, coordinate transformation equations are obtained as

$$r = f(r) = \begin{cases} r', & r' \geq b, \\ \dfrac{b^2}{r'}, & a \leq r' \leq b, \\ \left(\dfrac{b}{a}\right)^2 r', & b \leq r'. \end{cases}$$

Here, $e = \dfrac{b^2}{a}$, where $a$ and $b$ are core and outer complementary radius of AOSC respectively, and $e$ is outer boundary of it. So, based on transformation theory, mass density tensor and bulk modulus in the acoustic cloaking structure are obtained as $\rho'(\vec{r}')^{-1} = \dfrac{\Lambda \rho(\vec{r})^{-1} \Lambda^T}{\det \Lambda}$, $\kappa'(\vec{r}) = \det \Lambda \times \kappa(\vec{r})$, respectively, where $\Lambda$ is Jacobean transformation matrix with $\Lambda_{ij} = \dfrac{\partial x'_i}{\partial x_j}$ component. If $\rho_0$ and $\kappa_0$ are background material acoustic parameters, then

$$\rho' = \begin{bmatrix} -\rho_0 & 0 \\ 0 & -\rho_0 \end{bmatrix},$$

$$\kappa' = \dfrac{r^4}{b^4} \kappa_0,$$

are complementary material parameters, anti-image of object parameters embedded in complementary area are calculated as

$$\rho'_{anti-object} = \begin{bmatrix} -\rho_{object} & 0 \\ 0 & -\rho_{object} \end{bmatrix},$$

$$\kappa'_{anti-object} = \dfrac{r^4}{b^4} \kappa_{object},$$

And core material parameters are as follows

$$\rho'' = \begin{bmatrix} \rho_0 & 0 \\ 0 & \rho_0 \end{bmatrix},$$

$$\kappa'' = \dfrac{a^4}{b^4} \kappa_0,$$

Additionally, in this arrangement, the object and the AOSC are covered with EMPC to hide them from the electromagnetic point of view. The EMPC outer radius should not be larger than e, because the AOSC does not work out of this area. see Ref- [2] for the EMPC, coordinate transformations equations are

$$r' = c + \frac{d-c}{d} r,$$
$$\varphi' = \varphi,$$
$$z' = z.$$

EMPC permittivity ($\varepsilon'$) and permeability ($\mu'$) tensors have been calculated based on the theory of coordinate transformations as $\varepsilon' = \frac{\Lambda \varepsilon \Lambda^T}{\det \Lambda}$, $\mu' = \frac{\Lambda \mu \Lambda^T}{\det \Lambda}$. Finally, EMPC tensor components are derived as

$$\varepsilon_r = \mu_r = \frac{r-c}{r},$$
$$\varepsilon_\varphi = \mu_\varphi = \frac{r}{r-c},$$
$$\varepsilon_z = \mu_z = \frac{r-c}{r}\left(\frac{c}{e-c}\right)^2.$$

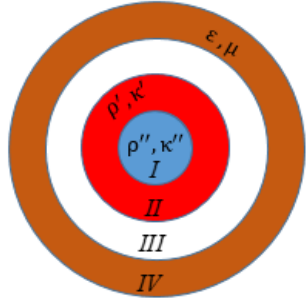

**Figure 7: Segmentation of blind arrangement areas.**

### Deaf cloak

As shown in Fig.8, in this case, an EMOSC for electromagnetic and a PAC for acoustic cloaking are used. Like the blind mode, Jacobin's matrix and coordinates transformations for cloaking structures are obtained. The procedures used in Ref- [2-3] have been used to obtain EMOSC and PAC parameters, respectively. Referring to Ref.[3] $r = 3 - 2r'$, $\varphi' = \varphi$, $z' = z$ are transformation equations in region II. Hence the permittivity and the permeability tensors in the complementary area have been obtained as

$$\varepsilon_r' = \mu_r' = 1 - \frac{3}{2r'},$$

$$\varepsilon_\varphi' = \mu_\varphi' = \frac{2r'}{2r' - 3},$$

$$\varepsilon_z' = \mu_z' = 2\frac{(2r' - 3)}{r'}.$$

Also $r'' = (a/c)r$, $\varphi'' = \varphi$, $z'' = z$ are transformation equations in region I. Then material parameters in this region have been derived as

$$\varepsilon_r'' = \mu_r'' = \varepsilon_r'' = \mu_r'' = 1, \varepsilon_z'' = \mu_z'' = \left(\frac{e}{a}\right)^2$$

by assuming the background to be air.

According to Ref-[2], PAC parameters are $\rho_r = \rho_0 \frac{r-c}{r}$ and $\rho_\varphi = \rho_0 \frac{r}{r-c}$ for mass density tensor and $\kappa = \kappa_0 \frac{r}{r-c}\left(\frac{d-c}{c}\right)^2$ for bulk modulus, which compress regions I, II and III into region IV, as to transformation coordinate theory.

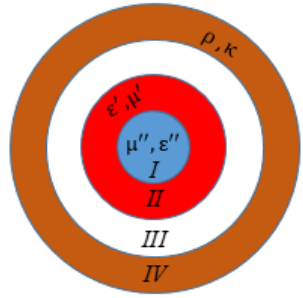

**Figure 8: Segmentation of deaf arrangement areas.**

**Result and discussion**

In this section, electromagnetic and acoustic waves scattering patterns have been shown for both blind and deaf modes. Acoustic wave scattering, according to Ref- [17], can be studied using the Maxwell equations. In fact, in Ref- [17], the acoustic propagation equations are mapped to Maxwell's equations with a substitution of their variables and the problem of the acoustic waves scattering turns into the electromagnetic waves scattering. The procedure presented in Ref- [17] is used to simulate the acoustic cloaking. On the other hand, one of the powerful methods for simulating Maxwell's equations is the Finite Difference Time Domain (FDTD). The dispersive FDTD method presented in Ref- [30] is used to simulate the scattering of acoustic and electromagnetic waves because the conventional FDTD simulation cannot correctly produce the materials applied in the invisible structures.

**Blind mode**

Referring to Table I, the parameters for AOSC and EMPC have provided. Furthermore, the propagation of acoustic and EM waves in the presence of blind structures have been simulated.

Table I: structural parameter of blind mode arrangement.

| Parameter | Value(cm) | Description |
|---|---|---|
| $a$ | 0.5 | AOSC Core radius |
| $b$ | 1 | AOSC complementary inner radius |
| $C$ | 1.6 | AOSC complementary outer radius |
| $d$ | 1.8 | EMPC inner radius |
| $e$ | 2 | EMPC outer radius |
| $\lambda_{EM}$ | 1 | Electromagnetic wave length |
| $\lambda_{Acoustic}$ | 1 | Acoustic wave length |

As shown in Fig.9, the EM waves, without penetrating inside the EMPC, will pass through it. Due to the design of the EMPC, waveform and phase are unchanged, so the objects inside the EMPC are invisible because there is not an electromagnetic link between the object in cloaking area and outside the construction. In Fig.10, waveform and phase of the scattering acoustic waves, the same as EM waves, are unaffected by the structure and an acoustic plane wave saves their shape after interacting with the structure, while the object receives acoustic signal. So, while the object acoustically and electromagnetically has been cloaked, it has not EMw connection, which means it is blind, but it can hear acoustic signals.

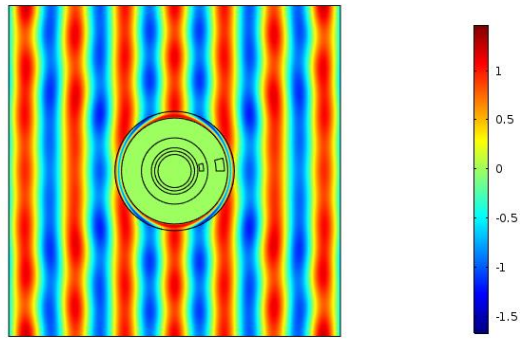

**Figure 9: EMW pattern interface with blind arrangement, without penetration inside the EMPC.**

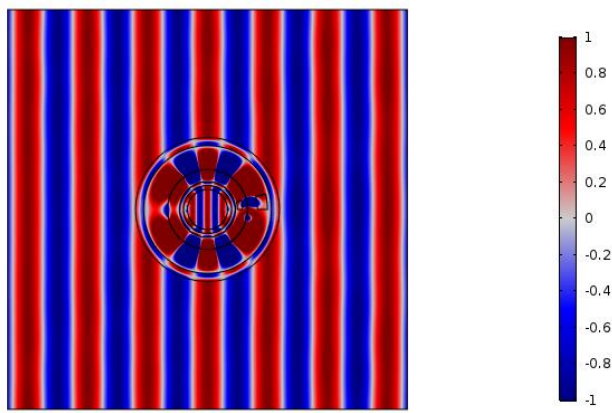

**Figure 10: acoustic wave propagation pattern and its interaction with cloaking structure.**

**Deaf mode**

The parameters used in transformations and geometric design are shown in Table II.

Table III : structural parameter of deaf mode arrangement.

| Parameter | Value (cm) | Description |
|---|---|---|
| $a$ | 0.5 | EMOSC core radius |
| $b$ | 1 | EMOSC complementary inner radius |
| $c$ | 1.6 | EMOSC complementary outer radius |
| $d$ | 1.8 | PAC inner radius |
| $e$ | 2 | PAC outer radius |
| $\lambda_{EM}$ | 1 | Electromagnetic wave length |
| $\lambda_{Acoustic}$ | 0.5 | Acoustic wave length |

The scattering of electromagnetic and acoustic waves can be seen in Figs.11 and 12, respectively. In Fig.11 electromagnetic invisibility and in Fig.12 acoustic cloaking can be seen. In this mode, sound waves do not have the ability to penetrate into the structure and do not interact with the inner structure. Therefore, the cloaking is called deaf. However, in this mode EMw link to the object has turned on.

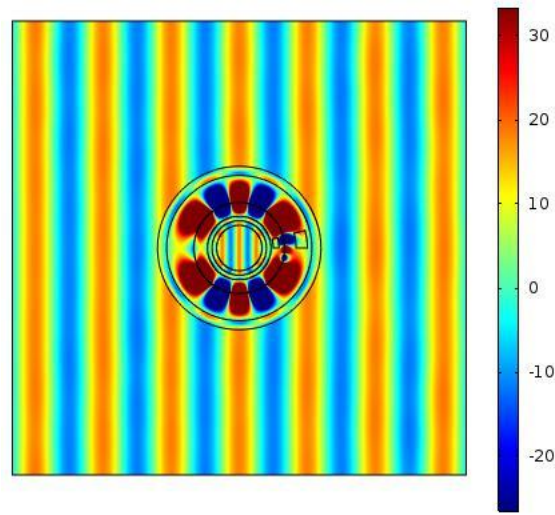

**Figure 11: EMW interaction with deaf mod arrangement.**

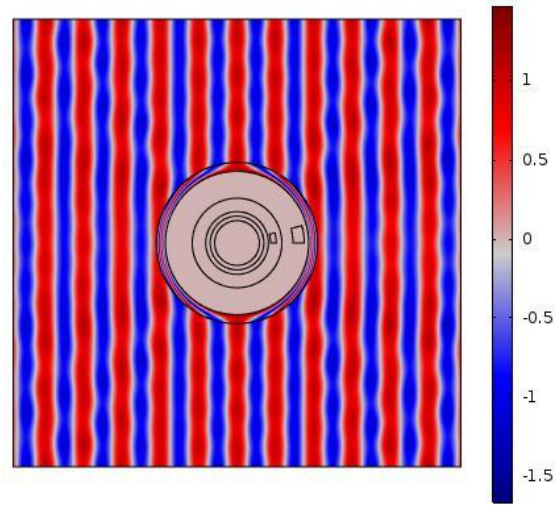

**Figure 12: acoustic wave interaction with deaf mode arrangement.**

## Conclusion

In this paper we evaluated the possibility of electromagnetic and acoustic double- cloaking of an object, with a combination of two common cloaking approaches for acoustic and electromagnetic cloaking. The results showed that the proposed method works for the double-cloaking. This method also has the ability to be generalized to any method of cloaking and it is possible to create other dual cloaking such as electromagnetic-elastic, sound-wave, and so on.

**References**

1. J. B. Pendry, D. Schurig, and D. R. Smith, Science **312**, 1780 (2006).

2. S. A. Cummer, B. I. Popa, D. Schurig, D. R. Smith, and J. Pendry, Physical Review E - Statistical, Nonlinear, and Soft Matter Physics **74**, 1 (2006).

3. M. Gharghi, C. Gladden, T. Zentgraf, Y. Liu, X. Yin, J. Valentine, and X. Zhang, Nano Letters **11**, 2825 (2011).

4. J. Li and J. B. Pendry, Physical Review Letters **101**, 1 (2008).

5. Y. Lai, H. Chen, Z. Q. Zhang, and C. T. Chan, Physical Review Letters **102**, 1 (2009).

6. B. N. B. Kundtz, D. R. Smith, and J. B. Pendry, Proceedings of the IEEE **99**, 1623 (2011).

7. H. F. Ma and T. J. Cui, Nature Communications **1**, 124 (2010).

8. D. H. Kwon and D. H. Werner, New Journal of Physics **10**, (2008).

9. J. B. Pendry, Physical Review Letters **85**, 3966 (2000).

10. J. P. Turpin, A. T. Massoud, Z. H. Jiang, P. L. Werner, and D. H. Werner, Optics Express **312**, 1777 (2006).

11. D. A. Roberts, N. Kundtz, and D. R. Smith, Optics Express **17**, 16535 (2009).

12. J. B. Pendry, A. I. Fernández-Domínguez, Y. Luo, and R. Zhao, Nature Physics **9**, 518 (2013).

13. L. Xu and H. Chen, Nature Photonics **9**, 15 (2014).

14. X. Zhu, B. Liang, W. Kan, X. Zou, and J. Cheng, Physical Review Letters **106**, 1 (2011).

15. S. A. Cummer, B. I. Popa, D. Schurig, D. R. Smith, J. Pendry, M. Rahm, and A. Starr, Physical Review Letters **100**, 1 (2008).

16. B. I. Popa and S. A. Cummer, Physical Review B - Condensed Matter and Materials Physics **83**, 1 (2011).

17. H. Chen and C. T. Chan, Journal of Physics D: Applied Physics **43**, (2010).

18. N. Stenger, M. Wilhelm, and M. Wegener, Physical Review Letters **108**, 1 (2012).

19. R. Schittny, M. Kadic, S. Guenneau, and M. Wegener, Physical Review Letters **110**, 1 (2013).

20. S. Brûlé, E. H. Javelaud, S. Enoch, and S. Guenneau, Physical Review Letters **112**, 1 (2013).

21. Y. A. Urzhumov and D. R. Smith, Physical Review Letters **107**, 2 (2011).

22. F. Monticone and A. Alù, Physical Review X **3**, 1 (2014).

23. Y. H. Su, J. W. Shi, D. H. Liu, and G. J. Yang, Chinese Physics Letters **27**, 1 (2010).

24. S. R. Sklan, X. Bai, B. Li, and X. Zhang, Scientific Reports **6**, 1 (2016).

25. X. Shi, F. Gao, X. Lin, and B. Zhang, Scientific Reports **5**, 1 (2015).

26. B. Zhang, T. Chan, and B. I. Wu, Physical Review Letters **104**, 1 (2010).

27. Y. Ma, Y. Liu, M. Raza, Y. Wang, and S. He, Physical Review Letters **113**, 1 (2014).

28. Y. Yang, H. Wang, F. Yu, Z. Xu, and H. Chen, Scientific Reports **6**, 1 (2016).

29. Y. Zhao, C. Argyropoulos, and Y. Hao, **16**, 6717 (2008).